# AB-Type Dual-Phase High-Entropy Alloys as Negative Electrode of Ni-MH Batteries: Impact of Interphases on Electrochemical Performance

Shivam Dangwal[1,2], Yongtao Li[3,*] and Kaveh Edalati[1,2,*]

[1] WPI, International Institute for Carbon-Neutral Energy Research (WPI-I2CNER), Kyushu University, Fukuoka 819-0395, Japan
[2] Department of Automotive Science, Graduate School of Integrated Frontier Sciences, Kyushu University, Fukuoka 819-0395, Japan
[3] School of Materials Science and Engineering, Anhui University of Technology, Maanshan, China

High-entropy alloys (HEAs) and their corresponding high-entropy hydrides are new potential candidates for negative electrode materials of nickel-metal hydride (Ni-MH) batteries. This study investigates the cyclic electrochemical hydrogen storage performance of two AB-type HEAs (A: hydride-forming elements, B: non-hydride-forming elements) in Ni-MH batteries. TiV$_2$ZrCrMnFeNi with a dual-phase structure shows a fast activation and a low charge transfer impedance with a discharge capacity of 150 mAhg$^{-1}$, while TiV$_{1.5}$Zr$_{1.5}$CrMnFeNi with a single phase shows a slow activation and a capacity of only 60 mAhg$^{-1}$. The better electrochemical performance of TiV$_2$ZrCrMnFeNi was attributed to its higher vanadium/zirconium ratio and abundant interphase boundaries, which act as hydrogen paths and heterogeneous hydride nucleation sites. These results suggest the high potential of dual-phase HEAs as new active electrode materials for Ni-MH batteries.
***Keywords:*** Rechargeable battery; Hydrogen storage materials; Metal hydrides; Multi-principal element alloys (MPEAs); Anode materials

*Corresponding authors
  Kaveh Edalati (E-mail: kaveh.edalati@kyudai.jp; Tel/Fax: +81 92 802 6744)
  Yongtao Li (E-mail: liyongtao@ahut.edu.cn; Tel: +86-555-231-1570)

**Introduction**

Decarbonization of the transportation sector plays a crucial role in the pursuit of human civilization towards a low $CO_2$ emission society [1]. A major effort in the decarbonization of the transportation sector is the fast development of electric and hybrid electric vehicles, in which energy is stored in rechargeable batteries [1,2]. Various kinds of batteries can be used in an electric vehicle such as lead-acid (Pb-acid), nickel-cadmium (Ni-Cd), lithium-ion (Li-ion), sodium-nickel-chloride (Na-Ni-Cl) and nickel-metal hydride (Ni-MH) batteries [3]. Each of these batteries has some advantages and disadvantages. For example, Pb-acid batteries have a low energy/weight ratio; Ni-Cd batteries have the highest lifespan but the use of cadmium hurts the environment; Li-ion batteries have a high energy density/weight ratio but lithium is expensive and its resources are limited; and Na-Ni-Cl batteries are characterized by high stored energy density but are compromised by operational safety [3]. Ni-MH batteries exhibit high energy density, long lifespan, a broad range of operating temperatures and do not contain poisonous elements [2,4]. Large-scale production of Ni-MH batteries started in the 1990s and since then there have been significant developments in this industry [4]. These developments resulted in the increase of gravimetric energy density (from 54 to 110 Wh/kg), volumetric energy density (from 190 to 490 Wh/L) and power (from 200 to 1200 W/kg) for commercial applications [4,5]. For example, as of 2017, 85% of hybrid electric vehicles used Ni-MH batteries [4,5]. However, the use of Ni-MH batteries for both stationary and mobile applications has been reduced in recent years because their energy/weight ratio is less than Li-ion batteries [2]. Ni-MH batteries are still a preferred choice for many energy storage applications, provided that their energy/weight ratio or discharge capacity can be improved [6].

Ni-MH batteries usually have a negative electrode of metal hydride such as $LaNi_5H_6$ which stores hydrogen in a solid-state form, a positive electrode of Ni hydroxide, and an electrolyte of potassium hydroxide [2,6]. The energy in Ni-MH batteries is stored/consumed by the movement of hydroxide ions between the two electrodes, as shown in the following half-cell reactions [2,6].

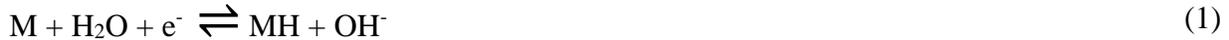

$$M + H_2O + e^- \rightleftharpoons MH + OH^- \qquad (1)$$

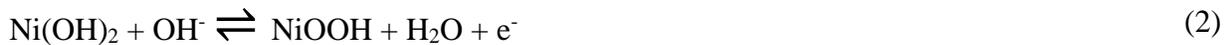

$$Ni(OH)_2 + OH^- \rightleftharpoons NiOOH + H_2O + e^- \qquad (2)$$

The key point in enhancing the discharge capacity of Ni-MH batteries is increasing the hydrogen absorption/desorption capacity of metal hydride in the negative electrode [7,8]. For Ni-MH batteries, the absorption and desorption of hydrogen in metal hydrides should occur reversibly at ambient temperature with fast kinetics and without the need for an activation process, but there are quite limited hydrogen storage materials and metal hydrides that can satisfy these conditions [7,8].

Various types of hydrogen storage alloys are used as electrode material for Ni-MH batteries which are either Mg-based, Ti-V-based, RE-Mg-Ni-based (RE: rare-earth elements), $AB_5$-type (A: hydride-forming elements; B: non-hydride-forming elements) or $AB_2$-type. Mg-based alloys containing long-period stacking order (LPSO) structures exhibit elevated hydrogen storage capacity and fast kinetics [9]. These LPSO structures can also show a high capacity for Ni-MH battery applications [10]. V-based alloys generally have high hydrogen absorption capacities, but their reversibility is usually poor due to the formation of stable hydrides. The synergistic use of vanadium and titanium provides better reversible electrochemical properties [11]. RE-Mg-Ni-based alloys, which are considered LPSO structures with stackings of $AB_5$ and $AB_2$, show high discharge capacities with a rather low cyclability [4,12]. Among all these materials, $AB_5$-type [13-16] and $AB_2$-type [17-20] alloys are still the main materials studied for Ni-MH electrodes [7]. For $AB_5$-type materials, $LaNi_5$ is the most popular compound, while partial substitution of nickel by cobalt can improve the cycling stability of LaNi$_5$ [21]. Although AB$_5$-type alloys have demonstrated high durability for Ni-MH battery applications, enhancing their capacity remains limited due to their crystal structures [4]. AB$_2$-type alloys are characterized by C14 Laves phase and BCC-type structures, both of which exhibit a large number of interstitial tetrahedral sites for hydrogen uptake [22]. It was reported that tuning the ratio of the C14 to BCC phases can enhance the activation, hydrogen storage capacity and discharge ability of AB$_2$-type alloys for Ni-MH battery applications [4,23,24]. Despite the significant progress of these materials for the negative electrode of Ni-MH batteries, there are still demands to discover new compounds with higher hydrogen storage capacities at room temperature.

A new class of materials known as high-entropy alloys (HEAs), containing at least five principal elements and a configurational entropy greater than 1.5$R$ ($R$: gas constant), has recently demonstrated high potential for hydrogen storage [25,26]. Reversible hydrogen storage with good kinetics and good activation was reported in some of these materials [27-29], but there is only a single study that used these materials for Ni-MH batteries [30]. In this earlier study, an AB$_2$-type HEA TiZrCrMnFeNi exhibited a discharge capacity of about 50 mAhg$^{-1}$ as the negative electrode of Ni-MH batteries [30]. Such a discharge capacity is significantly lower than the capacity of rare-earth-based AB$_5$-type electrode materials, and thus, new strategies should be employed to enhance the performance of HEAs for Ni-MH batteries.

In this study, two AB-type HEAs (TiV$_2$ZrCrMnFeNi and TiV$_{1.5}$Zr$_{1.5}$CrMnFeNi) were examined as negative electrode materials of Ni-MH batteries. These materials were designed by adding vanadium to previously developed TiZrCrMnFeNi alloy by considering two facts. (i) It is well-known that Ti-V-Cr/Mn-based materials can exhibit good hydrogen storage properties at room temperature under a gaseous hydrogen atmosphere [31,32]. (ii) The alloys based on Zr-V-Ni were reported to show good electrochemical hydrogen storage properties [33-35]. TiV$_{1.5}$Zr$_{1.5}$CrMnFeNi was synthesized as a single-phase alloy with C14 Laves phase structure; however, TiV$_2$ZrCrMnFeNi was intentionally synthesized as a dual-phase alloy with BCC and C14 phases because interphase boundaries are believed to be effective for activation of high-entropy hydrogen storage compounds [28,29]. An improvement in discharge capacity to 150 mAhg$^{-1}$ was achieved for the dual-phase AB-type HEA due to the contributions of V-rich composition and interphases.

**Experimental Procedure**

Arc melting was used to synthesize the HEAs TiV$_2$ZrCrMnFeNi and TiV$_{1.5}$Zr$_{1.5}$CrMnFeNi using metallic pieces of high-purity titanium (99.9%), vanadium (99.7%), zirconium (99.2%), chromium (99.99%), manganese (99.9%), iron (99.9%) and nickel (9.99%) on a water-cooled copper mold. The chamber after the initial evacuation was filled with argon under a pressure of 0.06 MPa to maintain an inert atmosphere for arc melting. The samples were flipped and remelted seven times to achieve good homogeneity.

X-ray diffraction (XRD) was used for the characterization of the crystal structure of HEAs using Cu K$\alpha$ at an acceleration voltage of 45 kV and a filament current of 200 mA. The crystal structure of hydride was examined by XRD after hydrogenation of samples in a Sivert-type machine, as described in detail in an earlier publication [27]. The phases and lattice parameters in the crystal structure were confirmed by analyzing the XRD profiles using PDXL software.

The examination of the microstructure was carried out using a scanning electron microscope (SEM) at an acceleration voltage of 15 kV. For SEM, the samples were cut into discs using a wire-cutting machine followed by mechanical grinding and fine polishing to achieve a mirror-like surface. The phase fraction in SEM was determined using an image segmentation tool in the MATLAB software. The chemical and phase analyses of the microstructure were carried out using energy-dispersive X-ray spectroscopy (EDS) and electron backscatter diffraction (EBSD).

The oxidation states of elements on the surface of samples were examined by X-ray photoelectron spectroscopy (XPS) using the Al Kα radiation. The samples prepared for SEM were employed for XPS analysis as well.

The HEAs were used as negative electrode material for Ni-MH batterie, as schematically shown in Fig. 1. The electrochemical performance of HEAs was assessed at room temperature by making a three-electrode electrochemical cell equipped with a potentiostat-galvanostat device. The negative electrode was created by crushing the HEA to particle sizes of <100 µm followed by mixing it with pure copper (99.999%) powder of particle sizes less than 45 µm. This mixture was further used to fabricate a disc with a diameter of 13 mm and a thickness of 0.5 mm by using a manual press under a pressure of 10 MPa. The disc was placed within a copper holder, as described in detail in an earlier publication [300], and it was used as the negative working electrode in 6M KOH electrolyte with $Ni(OH)_2$/NiOOH as the positive electrode and Hg/HgO as the reference electrode. The cell underwent a charging phase at 100 mA/g for 6 h and a discharging phase at 50 mA/g to a cutoff potential of -0.6 V vs. Hg/HgO. This charging-discharging process was repeated 50 times to examine the cyclic stability.

Electrochemical impedance spectroscopy (EIS) was conducted on negative electrodes of Ni-MH batteries using a CHI 760E electrochemical workstation, which employed an AC voltage of 5 mV concerning the open-circuit potential. The scanning frequency for EIS was in the range of 0.1 to $10^6$ Hz.

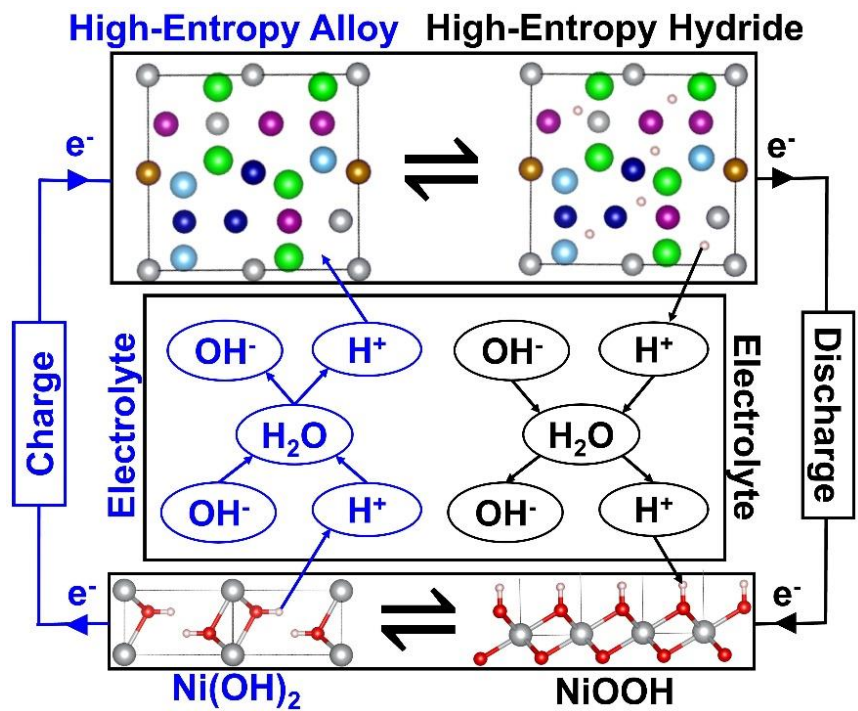

Figure 1. Illustration of reaction in the negative and positive electrode of Ni-MH batteries with high-entropy alloys as negative electrode materials.

**Results**

The crystal structure analysis of the HEAs by XRD is shown in Fig. 2(a). TiV$_2$ZrCrMnFeNi shows the presence of two phases: C14 Laves phase (P63/mmc with $a = b = 0.494$ nm and $c = 0.813$ nm) and BCC phase (Im-3m with $a = b = c = 0.295$ nm). For the HEA TiV$_{1.5}$Zr$_{1.5}$CrMnFeNi, the XRD analysis shows the presence of a single-phase C14 structure ($a = b = 0.499$ nm and $c = 0.817$ nm). The presence of the C14 Laves phase is a positive issue because this phase is considered a suitable phase for hydrogen storage in HEAs [36-39]. Moreover, the presence of the second phase of BCC should be due to the high vanadium amount in TiV$_2$ZrCrMnFeNi [40]. This is also a positive issue because dual-phase structures can show enhanced electrochemical properties [2,8,33,41] and interphases can activate the material for hydrogenation [28,29].

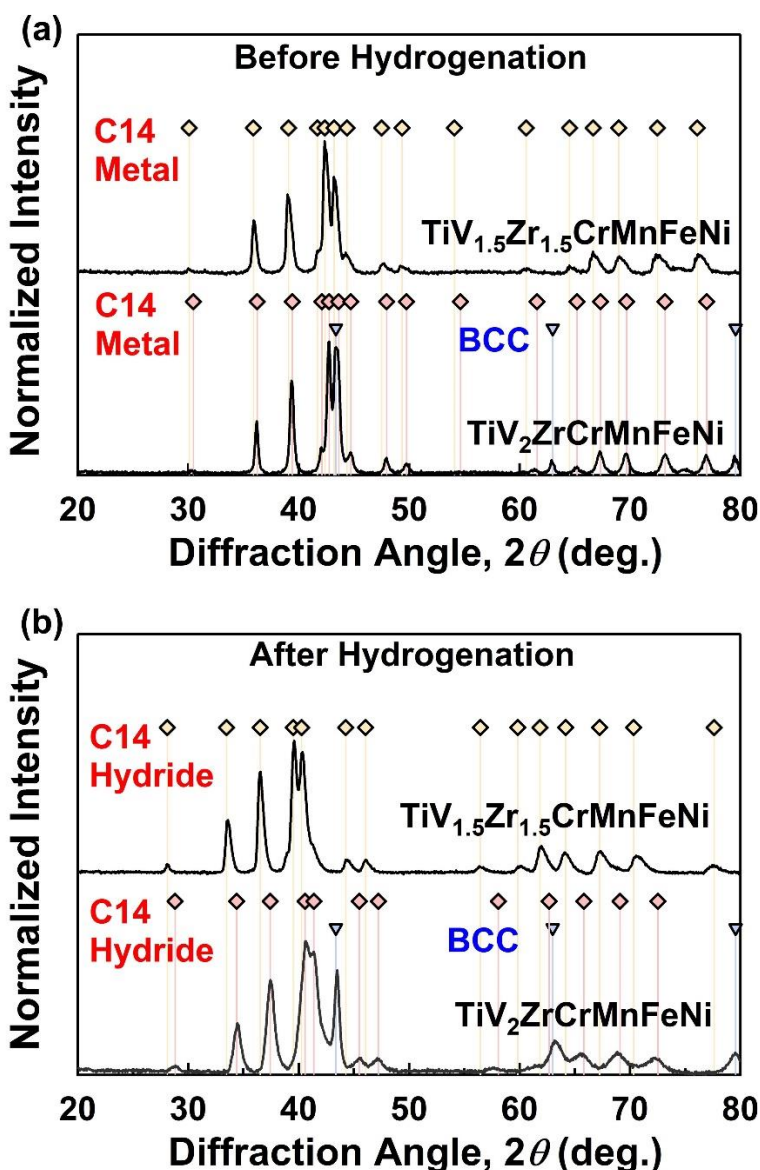

Figure 2. Transformation from high-entropy alloys with C14 Laves phase to high-entropy hydrides with C14 Laves phase. XRD profile (a) before and (b) after hydrogenation for TiV$_2$ZrCrMnFeNi and TiV$_{1.5}$Zr$_{1.5}$CrMnFeNi.

The XRD analysis after the hydrogenation of both HEAs is shown in Fig. 2(b). There is a peak shift for the C14 Laves phase after hydrogenation, indicating expansion of lattice due to Laves hydride formation. For TiV$_2$ZrCrMnFeNi, there is a 15% expansion in the C14 lattice with new lattice parameters ($a = b = 0.519$ nm and $c = 0.849$ nm), while the peaks of BCC remain unchanged. For TiV$_{1.5}$Zr$_{1.5}$CrMnFeNi, there is a 21% expansion in the C14 lattice with larger lattice parameters ($a = b = 0.532$ nm and $c = 0.873$ nm). The current results are reasonable because it was reported that HEAs with the C14 structure usually store hydrogen in the form of an expanded C14 phase [26].

The microstructure examined by SEM for TiV$_2$ZrCrMnFeNi is shown in Fig. 3(a), implying the presence of two phases with bright and dark contrasts. EBSD analysis illustrated in Fig. 3(b) confirms the presence of the Laves phase (bright contrast in SEM image) and the BCC phase (dark contrast in SEM image with ~20 vol%) in agreement with the XRD analysis. EDS elemental mappings in Fig. 3(c) verify that the BCC phase is rich in vanadium and chromium, as described quantitatively in Table 1. The presence of large fractions of C14/BCC interphases in Fig. 3 is expected to facilitate electrochemical reactions [4,22,23] and hydrogen storage [28,29].

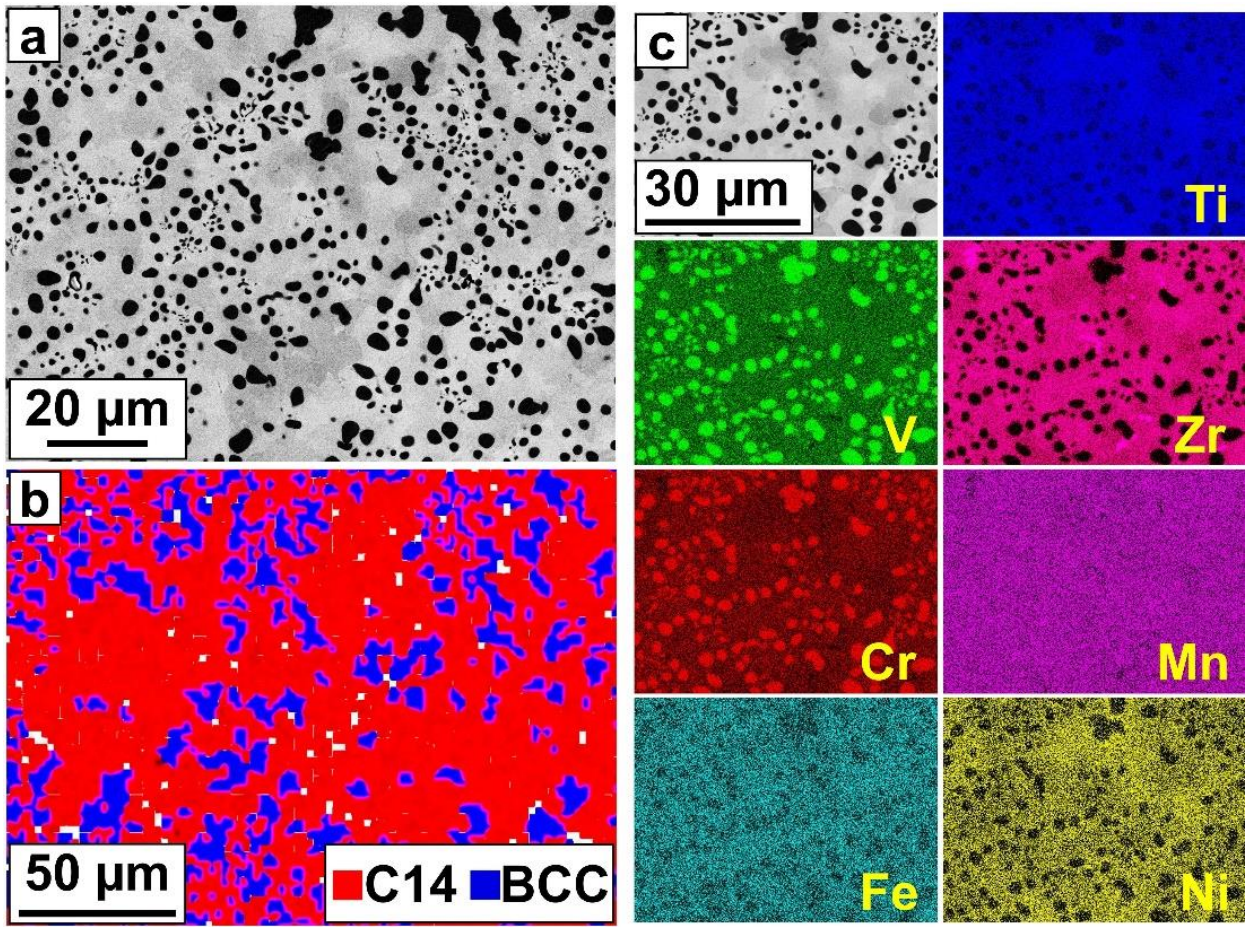

Figure 3. Formation of dual C14 and BCC phases in high-entropy alloy TiV$_2$ZrCrMnFeNi. (a) SEM micrograph, (b) EBSD phase map with a beam step size of 1 µm, and (c) SEM micrograph and corresponding EDS elemental mappings.

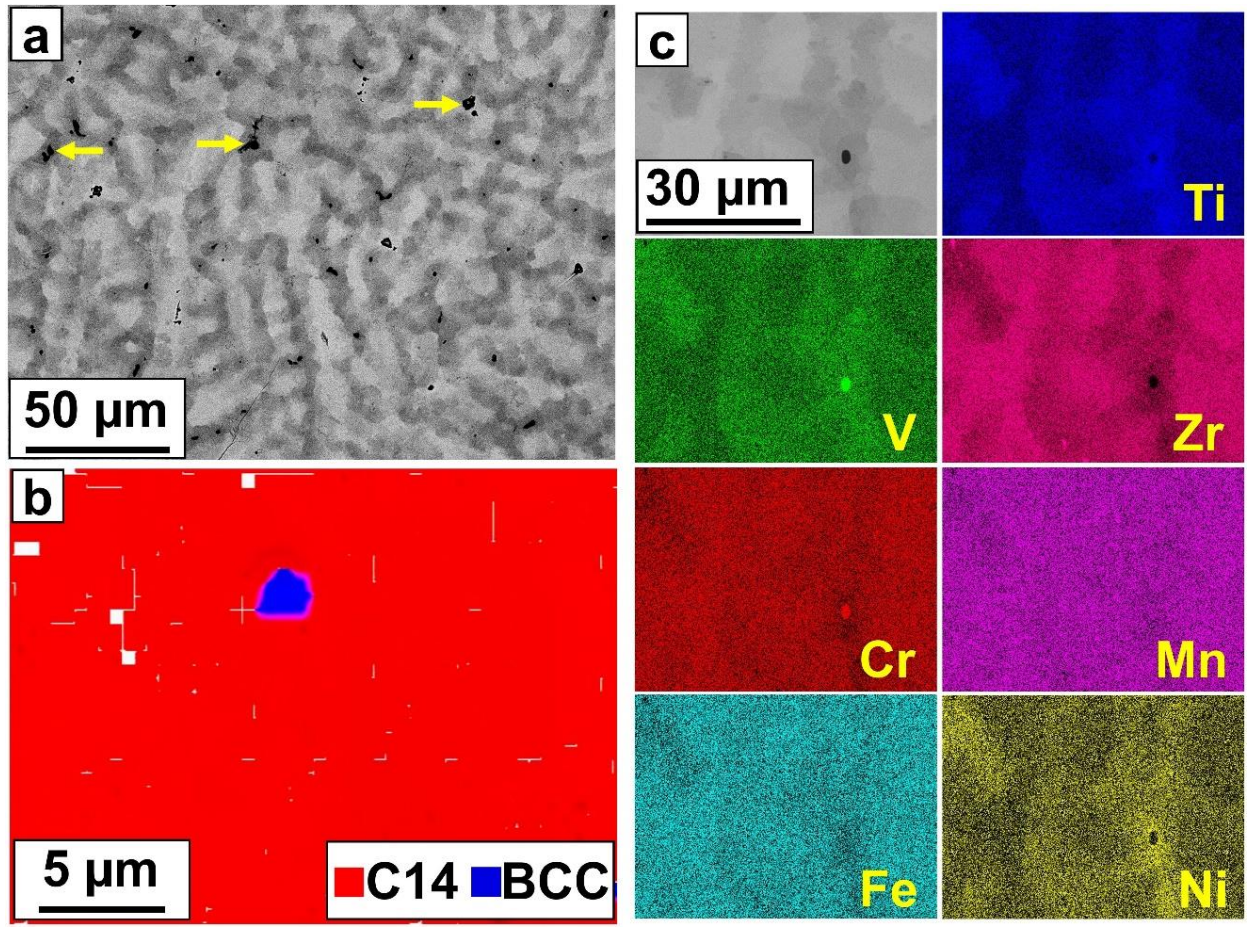

Figure 4. Formation of almost single-phase C14 structure with minor amounts of BCC phase in high-entropy alloy TiV$_{1.5}$Zr$_{1.5}$CrMnFeNi. (a) SEM micrograph, (b) EBSD phase map with a beam step size of 1 µm, and (c) SEM micrograph and corresponding EDS elemental mappings.

Table 1. Chemical compositions of high-entropy alloys TiV$_2$ZrCrMnFeNi and TiV$_{1.5}$Zr$_{1.5}$CrMnFeNi.

|  | TiV$_2$ZrCrMnFeNi | | | TiV$_{1.5}$Zr$_{1.5}$CrMnFeNi | | |
|---|---|---|---|---|---|---|
| **Element** | **Nominal** | **Overall** | **C14** | **BCC** | **Nominal** | **Overall** | **C14** |
| **Ti (at%)** | 12.5 | 12.8 | 13.5 | 5.9 | 12.5 | 12.6 | 12.8 |
| **V (at%)** | 25 | 25.6 | 18.8 | 51.6 | 18.75 | 19.0 | 18.9 |
| **Zr (at%)** | 12.5 | 13.4 | 17.9 | 0.2 | 18.75 | 19.8 | 19.5 |
| **Cr (at%)** | 12.5 | 12.8 | 11.3 | 21.7 | 12.5 | 12.5 | 12.6 |
| **Mn (at%)** | 12.5 | 10.2 | 10.8 | 9.4 | 12.5 | 11.5 | 11.2 |
| **Fe (at%)** | 12.5 | 12.7 | 14.3 | 8.5 | 12.5 | 12.3 | 12.3 |
| **Ni (at%)** | 12.5 | 12.5 | 13.4 | 2.7 | 12.5 | 12.3 | 12.7 |

The microstructure examined by SEM for TiV$_{1.5}$Zr$_{1.5}$CrMnFeNi is shown in Fig. 4(a). The existence of a compositional gradient is observed in the C14 phase by the image contrast in the SEM micrograph, but the BCC phase is detected only in limited regions indicated by arrows. EBSD analysis in Fig. 4(b) confirms the presence of the C14 Laves phase with a minor fraction of the BCC phase (less than 0.5 vol%). EDS elemental mappings shown in Fig. 4(c) imply the presence of a compositional gradient in the Laves phase with a small fraction of the V-Cr-rich

BCC phase. The quantitative EDS analysis for this alloy is given in Table 1, indicating a high amount of zirconium in C14 as a stabilizer of the Laves phase [42].

The XPS analysis taken from the surfaces of both HEAs is shown in Fig. 5, in which peak positions reported in the literature are also included [43,44]. An inspection of XPS results indicates three important points. (i) All elements except nickel are in fully oxidized states ($Ti^{4+}$, $V^{5+}$, $Zr^{4+}$, $Cr^{3+}$, $Mn^{4+}$ and $Fe^{3+}$) and there are no appreciable differences between XPS profiles of these elements for the two alloys. (ii) Nickel is detected in both an oxidized state ($Ni^{2+}$) and a non-oxidized metallic state in both alloys. (iii) The XPS profiles for oxygen in Fig 5(h) indicate that the surface is covered not only by the oxides of elements but also by hydroxides. It is concluded that the surface of two alloys is reasonably similar, while the presence of metallic nickel on the surface may contribute to catalytic hydrogen dissociation [45].

The discharge potential versus discharge capacity is shown in Fig. 6(a) for both HEAs after 20 cycles. $TiV_2ZrCrMnFeNi$ shows a discharge capacity of around 160 mAhg$^{-1}$ while $TiV_{1.5}Zr_{1.5}CrMnFeNi$ shows a lower capacity of only 60 mAhg$^{-1}$. A clear plateau is not observed in both the samples possibly due to compositional gradients [46]. The discharge capacity of HEAs as a function of cycle number is summarized in Fig. 6(b). $TiV_2ZrCrMnFeNi$ exhibits a good cycling performance with a discharge capacity of 150 mAhg$^{-1}$ after 50 cycles. The discharge capacity of $TiV_{1.5}Zr_{1.5}CrMnFeNi$ is almost 0 mAhg$^{-1}$ after the first two cycles and then gradually increases to reach a maximum capacity of 60 mAhg$^{-1}$ after 50 cycles indicating a slow activation. Contrary to single-phase $TiV_{1.5}Zr_{1.5}CrMnFeNi$ with slow electrochemical activation, dual-phase $TiV_2ZrCrMnFeNi$ shows a rather fast activation and reaches a discharge capacity of 60 mAhg$^{-1}$ only in the second cycle. These results confirm the significance of a second phase on electrochemical activation of HEAs for hydrogenation, which is in line with the reports on conventional electrode materials [2,8].

EIS plots of imaginary impedance versus real impedance are shown in Fig. 7 for $TiV_2ZrCrMnFeNi$ and $TiV_{1.5}Zr_{1.5}CrMnFeNi$. Semicircular curves, whose radius indicates the charge transfer impedance of Ni-MH battery, are observed for both alloys. This charge transfer impedance mainly depends on the interface reaction resistance between the negative electrode and the electrolyte. For dual-phase $TiV_2ZrCrMnFeNi$ alloy, the semicircle radius is smaller, which indicates that its charge transfer impedance is lower and its interface reaction speed is faster compared to the single-phase $TiV_{1.5}Zr_{1.5}CrMnFeNi$ alloy. In good agreement with Fig. 6, it is then expected that the charge-discharge performance of $TiV_2ZrCrMnFeNi$ should be better than $TiV_{1.5}Zr_{1.5}CrMnFeNi$ [47].

**Discussion**

In this study, two HEAs with single-phase and dual-phase structures are used as negative electrode materials for Ni-MH batteries with a target to examine the effect of interphase boundaries on the discharge capacity of high-entropy electrode materials. Although the discharge capacity of these two HEAs is still smaller compared to rare-earth-based AB$_5$-type compounds [13-15] and other conventional anode materials [9-12], two important findings are derived from this work. First, compared to TiZrCrMnFeNi, used earlier as the first HEA for a similar battery application, the discharge capacity improves by a factor of 3 in this study due to the modification of composition and the addition of vanadium. Second, this study shows that the presence of a second phase and corresponding interphase boundaries are effective for fast electrochemical activation and enhancement of discharge capacity. The above two issues are discussed further here.

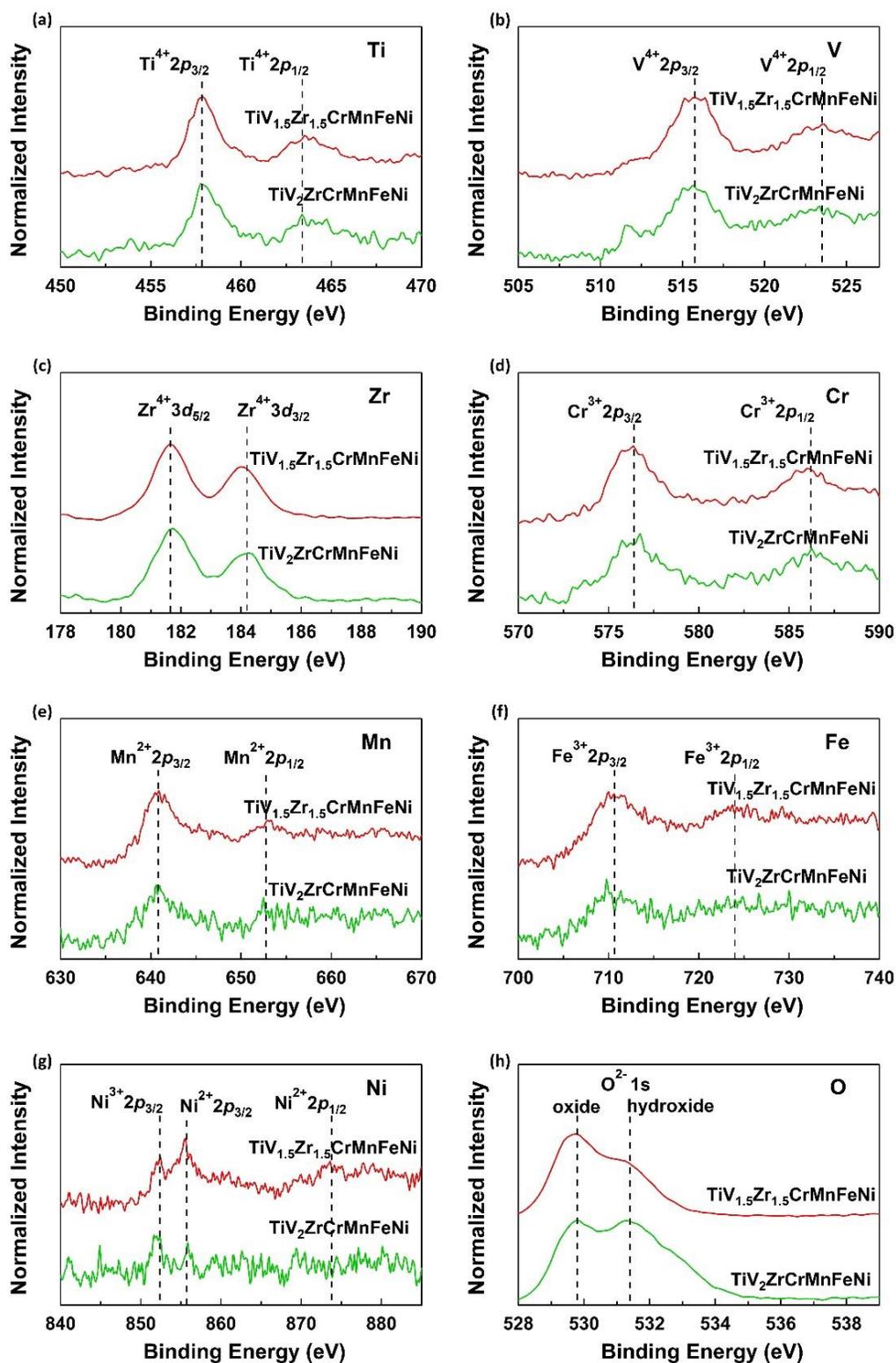

Figure 5. Presence of elements in oxidized states on surfaces high-entropy alloys. XPS profiles of (a) titanium, (b) vanadium, (c) zirconium, (d) chromium, (e) manganese, (f) iron, (g) nickel and (h) oxygen for TiV$_2$ZrCrMnFeNi and TiV$_{1.5}$Zr$_{1.5}$CrMnFeNi.

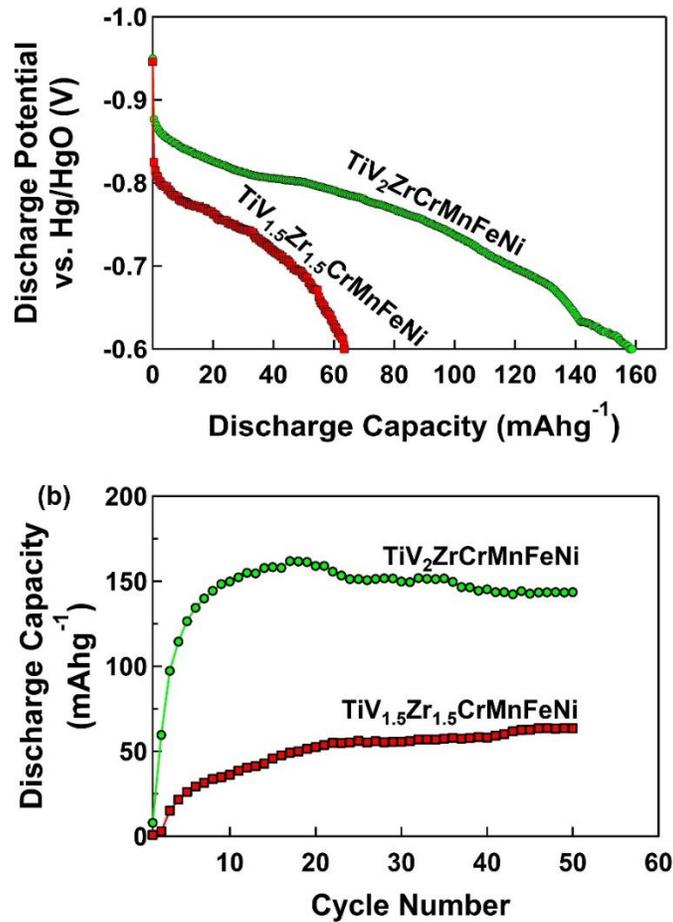

Figure 6. Good performance of dual-phase high-entropy alloys as negative electrode material for Ni-MH battery. (a) Discharge capacity versus discharge potential after 20 cycles, and (b) discharge capacity versus cycle number for TiV$_2$ZrCrMnFeNi and TiV$_{1.5}$Zr$_{1.5}$CrMnFeNi.

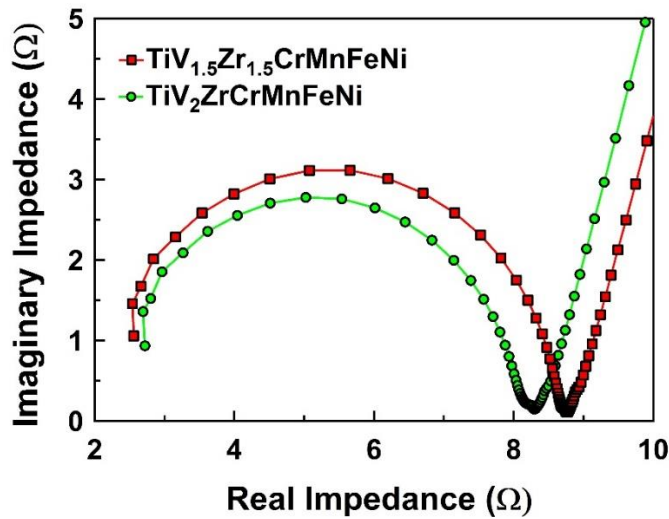

Figure 7. Lower charge transfer impedance of dual-phase high-entropy alloy for Ni-MH battery application. EIS profiles for TiV$_2$ZrCrMnFeNi and TiV$_{1.5}$Zr$_{1.5}$CrMnFeNi.

Regarding the first issue, it is noted that the storage media for AB-type TiV$_2$ZrCrMnFeNi with 150 mAhg$^{-1}$ capacity in this study and for AB$_2$-type TiZrCrMnFeNi with 50 mAhg$^{-1}$ capacity in an earlier study [30] are the C14 phases with almost similar lattice parameters. These alloys show almost similar room-temperature hydrogen storage performance in the presence of hydrogen gas, but their electrochemical behavior is quite different. Therefore, crystal structure and hydrogen storage capacity in a gaseous atmosphere are not the sole criteria for determining the discharge capacity, and composition should play a more critical role in electrochemical performance [4,5,21,22]. The enhancement of capacity in this study can be attributed to the following reasons. First, vanadium is an element having the smallest atomic radius amongst all other elements in these HEAs [48]. A high amount of the smallest size element creates a large atomic size mismatch which in turn forms more spacy interstitial sites for hydrogen to stay and increases the nucleation rate of the hydride phase [49]. Second, unlike the difficult activation of V-based hydrides in the air [31,32,50], a higher vanadium content does not cause significant difficulty in the electrochemical activation process due to the high solubility of vanadium in the alkaline medium of KOH [33]. Third, it is known that zirconium thermodynamically tends to form zirconium oxide in KOH which accordingly hinders the uptake of hydrogen by the electrode [35,41,51,52]. Since the amount of zirconium in TiV$_2$ZrCrMnFeNi is less than TiZrCrMnFeNi, its electrochemical performance can be partly justified by the oxidation of zirconium. It should be noted that despite this negative effect of zirconium, its addition to HEAs is essential to adjust the crystal structure and hydrogen binding energy for room-temperature hydrogen storage [37]. Earlier attempts to replace zirconium with other elements such as magnesium did not lead to room-temperature hydrogen storage [53].

Regarding the second, issue it should be noted that the Laves phase in TiV$_2$ZrCrMnFeNi and TiV$_{1.5}$Zr$_{1.5}$CrMnFeNi have similar crystal structures and almost similar compositions, but they show significantly different electrochemical performance. The main differences in the electrochemical performance are the easier activation, a higher discharge capacity (Fig. 6) and a lower charge transfer impedance (examined by EIS in Fig. 7) for TiV$_2$ZrCrMnFeNi. This difference can be hardly attributed to the surface conditions of the two alloys because XPS indicates significant similarity for these two alloys. The main reason for the difference can be due to the presence of a second phase within the C14 matrix. TiV$_2$ZrCrMnFeNi has a dual-phase structure with a large fraction of C14/BCC interphase boundaries, and it is believed that interphase boundaries not only act as hydrogen pathways for easy activation but also provide active sites for heterogeneous nucleation of the hydride phase [13,54]. Moreover, the presence of a second phase in the structure may act as a catalytic site for hydrogen uptake [2,41]. These effects of the second phase and interphases can be responsible for faster activity and a higher discharge capacity of TiV$_2$ZrCrMnFeNi. Similar to interphase boundaries, nanograin boundaries can enhance the hydrogen uptake; however, extra processing (e.g. ball milling or severe plastic deformation) is usually required after the initial synthesis for the formation of grain boundaries which is a demerit compared to the single-step generation of interphase boundaries [55-57].

In summary, although HEAs are a quite new family of electrode materials for Ni-MH batteries compared to traditional AB$_5$-type [13-16] and AB$_2$-type [17-20] alloys and their discharge capacities are still lower than the state-of-art [4,5], further research on these alloys, particularly dual-phase HEAs with interphase boundaries, is expected to lead to significant enhancement of their electrochemical performance with much higher discharge capacities. Such an expectation is due to the high flexibility in composition and phase adjustments of these entropy-stabilized materials for a particular application [25].

**Conclusion**

Two different kinds of AB-type HEAs were examined for their capacity as negative electrode materials of Ni-MH batteries. The dual-phase HEA TiV$_2$ZrCrMnFeNi with higher vanadium content showed faster electrochemical activation and higher discharge capacity compared to single-phase TiV$_{1.5}$Zr$_{1.5}$CrMnFeNi. The present study introduces V-rich dual-phase HEAs as potential candidates for Ni-MH battery applications.

**CRediT authorship contribution statement**

S.D., Y.L. and K.E.: Conceptualization, Methodology, Investigation, Writing – review & editing.

**Declaration of competing interest**

The authors declare that they have no known competing financial interests or personal relationships that could influence the work reported in this paper.

**Data availability**

Data will be made available on request.

**Acknowledgment**

The author S.D. thanks the MEXT, Japan for a scholarship. This work is supported in part by grants-in-aid for scientific research from the MEXT, Japan (JP19H05176 & JP21H00150).